\documentclass[cameraready]{Interspeech}

\usepackage[table]{xcolor}
\usepackage{colortbl}
\title{CodecFlow: Efficient Bandwidth Extension via Conditional Flow Matching in Neural Codec Latent Space}

\author[affiliation={1,2}, orcid=0009-0003-3927-6995]{Bowen}{Zhang*}
\author[affiliation={3}, orcid=0009-0008-2616-6590]{Junchuan}{Zhao*}
\author[affiliation={1}, orcid=0000-0001-7111-2008]{Ian}{McLoughlin}
\author[affiliation={3}, orcid=0000-0002-0123-1260]{Ye}{Wang}
\author[affiliation={2}, orcid=0000-0002-8740-634X]{A S}{Madhukumar}
\address{
    $^1$ Singapore Institute of Technology, Singapore \\
    $^2$ Nanyang Technological University, Singapore \\
    $^3$ National University of Singapore, Singapore
}

\email{bowen009@e.ntu.edu.sg, junchuan@u.nus.edu, 
ian.mcloughlin@singaporetech.edu.sg,
dcswangy@nus.edu.sg,
ASMadhukumar@ntu.edu.sg}


\usepackage{comment}
\usepackage{multirow}

\begin{document}

\maketitle

\begingroup
\renewcommand\thefootnote{*}
\footnotetext{These authors contributed equally to this work.}
\footnotetext{This work has been submitted to the IEEE for possible publication. Copyright may be transferred without notice, after which this version may no longer be accessible.}
\endgroup

\begin{abstract}
    Speech Bandwidth Extension improves clarity and intelligibility by restoring/inferring appropriate high-frequency content for low-bandwidth speech. Existing methods often rely on spectrogram or waveform modeling, which can incur higher computational cost and have limited high-frequency fidelity. Neural audio codecs offer compact latent representations that better preserve acoustic detail, yet accurately recovering high-resolution latent information remains challenging due to representation mismatch. We present CodecFlow, a neural codec-based BWE framework that performs efficient speech reconstruction in a compact latent space. CodecFlow employs a voicing-aware conditional flow converter on continuous codec embeddings and a structure-constrained residual vector quantizer to improve latent alignment stability. Optimized end-to-end, CodecFlow achieves strong spectral fidelity and enhanced perceptual quality on 8 kHz to 16 kHz and 44.1 kHz speech BWE tasks.
\end{abstract}

\section{Introduction}
Speech Bandwidth Extension (BWE) aims to reconstruct or infer appropriate high-frequency content from low-bandwidth speech signals to enhance perceptual quality and intelligibility. Early approaches primarily relied on signal-processing techniques, including source–filter models, mapping-based methods, and statistical estimation~\cite{chennoukh2001speech, unno2005robust, pulakka2011speech}. With the rise of deep learning, neural approaches have become the dominant paradigm. Initial DNN-based methods focused on either direct waveform reconstruction~\cite{ling2018waveform, birnbaum2019temporal} or the prediction of intermediate acoustic representations for high-fidelity resynthesis~\cite{pulakka2011bandwidth, kontio2007neural}.

Subsequently, Generative Adversarial Networks (GANs) gained prominence for improving perceptual quality by operating in the spectral domain or serving as neural vocoders. In the frequency domain, AERO~\cite{mandel2023aero} adopts an adversarial objective to predict full-band complex spectrograms. Similarly, mdctGAN~\cite{shuai2023mdctgan} performs adversarial learning on real-valued Modified Discrete Cosine Transform (MDCT) coefficients, enabling direct high-frequency synthesis without explicit phase reconstruction. Following a two-stage design, VoiceFixer~\cite{liu2022voicefixer} first restores clean mel-spectrograms using a ResUNet and then synthesizes waveforms via a pretrained TFGAN vocoder. To better model phase information, AP-BWE~\cite{lu2024towards} employs parallel lightweight CNN branches for amplitude and phase prediction, reinforced by multi-resolution discriminators for phase consistency. Beyond purely adversarial formulations, Fre-Painter~\cite{kim2024audio} frames bandwidth extension as a spectral inpainting problem and leverages a self-supervised masked autoencoder to learn contextual speech representations.

More recently, diffusion- and flow-based generative models have been explored to alleviate the training instability commonly observed in GANs. Nu-Wave2~\cite{han2022nu} proposed a diffusion-based model that integrates Short-Time Fourier Convolution and Bandwidth Spectral Feature Transform to perform arbitrary-scale audio super-resolution within a unified framework. To further improve inference efficiency, FlowHigh~\cite{yun2025flowhigh} adopts conditional flow matching to construct a direct probability path from noise to high-resolution mel-spectrograms, enabling efficient single-step generation while avoiding the multi-step sampling overhead typical of diffusion methods.

Concurrently, neural speech codecs have evolved beyond simple compression tools into powerful representation learners. Modern codecs can extract high-fidelity discrete and continuous latent representations that capture rich semantic and acoustic information. These representations have shown strong effectiveness in downstream generative tasks, including Text-to-Speech (TTS)~\cite{du2024cosyvoice, chen2025neural, ju2024naturalspeech, wangmaskgct, chen2025f5} and voice conversion (VC)~\cite{huang2024make, zhang2025vevo, DBLP:conf/interspeech/ZhaoWW25}.
 
Recent studies have explored latent representations from neural speech codecs, such as DAC~\cite{kumar2023high}, as intermediate features for BWE. For instance, \cite{liu2025neural} employ continuous DAC embeddings to directly predict clean discrete tokens from noisy inputs. In contrast, \cite{fang2025vector} adopt a generative strategy, conditioning a Discrete Denoising Diffusion Probabilistic Model \cite{austin2021structured} on these representations to reconstruct the target token sequence. Different from mel-spectrogram–based representations, which often discard phase information, and direct waveform modeling, which is computationally expensive for full-band generation, neural speech codecs provide compact latent representations that better preserve acoustic details while maintaining efficiency. However, accurately recovering high-resolution codec representations from low-bandwidth inputs remains challenging. Existing conversion-based methods, whether operating on discrete or continuous codec latents, frequently suffer from representation mismatch and information loss, which ultimately limits the perceptual quality of the reconstructed speech.

To address these challenges, we present \textbf{CodecFlow}, a neural codec-based bandwidth extension framework for efficient and high-fidelity speech reconstruction in a compact latent space. Our main contributions are as follows:
\begin{itemize}
    \item We propose a voicing-aware conditional flow-matching conversion model (FEC) on continuous codec latents with explicit V/UV conditioning, enabling more accurate recovery of high-frequency unvoiced components.

    \item We enhance a DAC-based codec with Structure-Constrained Residual Vector Quantization (SC-RVQ) to better align continuous embeddings and discrete tokens, improving representation stability for bandwidth extension.
    
    \item We fine-tune the entire framework end-to-end to jointly optimize codec representation learning and flow-based reconstruction for coherent low-to-high bandwidth eextension.

\end{itemize}

We evaluate CodecFlow against representative baselines using objective and perceptual metrics, demonstrating comparable or superior performance at both 16 kHz and 44.1 kHz with reduced spectral artifacts and improved perceptual quality.

\section{Method}
\begin{figure*}[t]
\centering
\includegraphics[width=\textwidth]{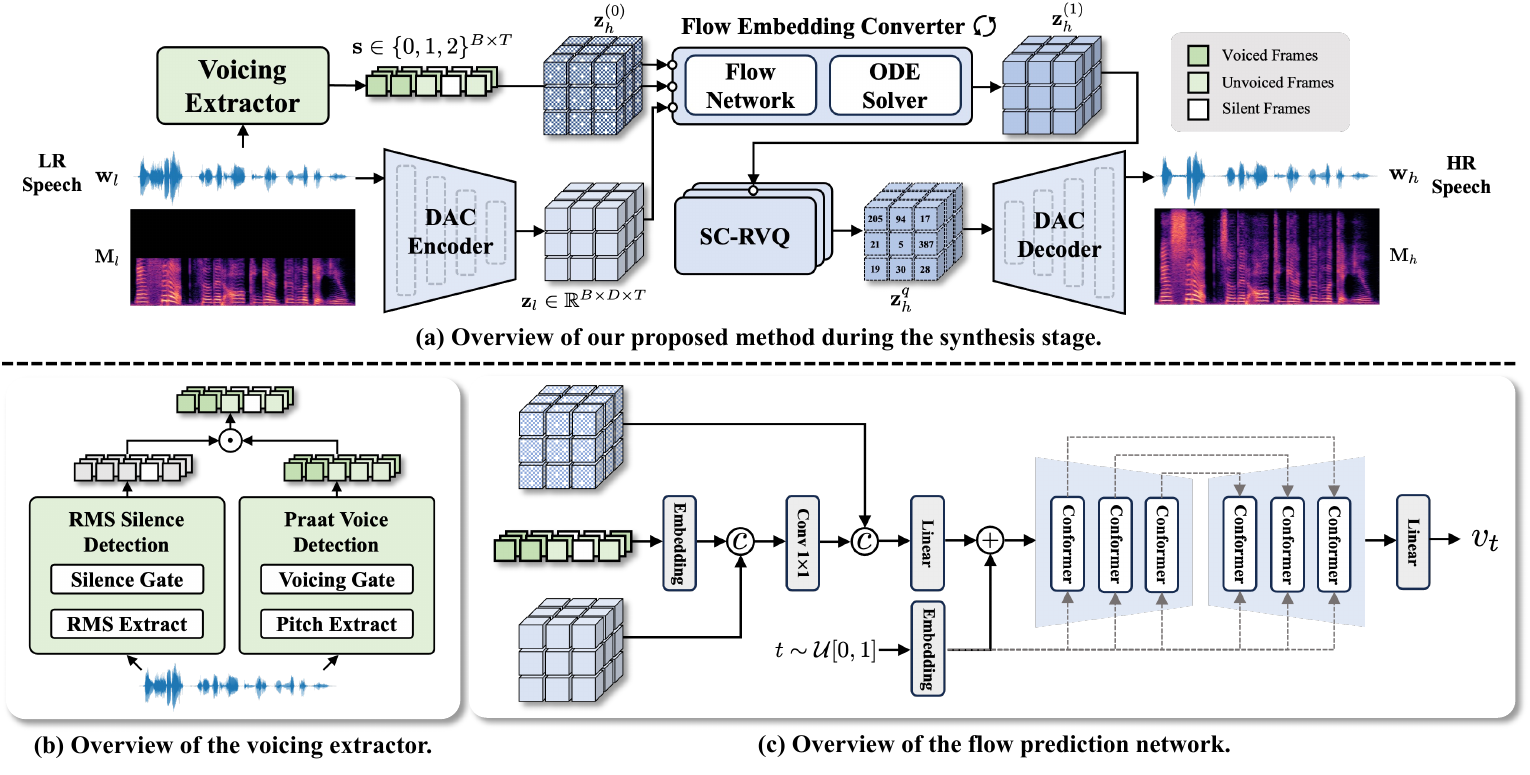}
\caption{An overview of the proposed CodecFlow framework, including (a) the overall model pipeline, (b) the architecture of the voicing extractor, and (c) the architecture of the flow prediction network from the flow embedding converter (FEC).}
\label{fig:overall}
\end{figure*}

\subsection{Overall Architecture}

\begin{figure}[t]
\centering
\includegraphics[width=0.5\textwidth]{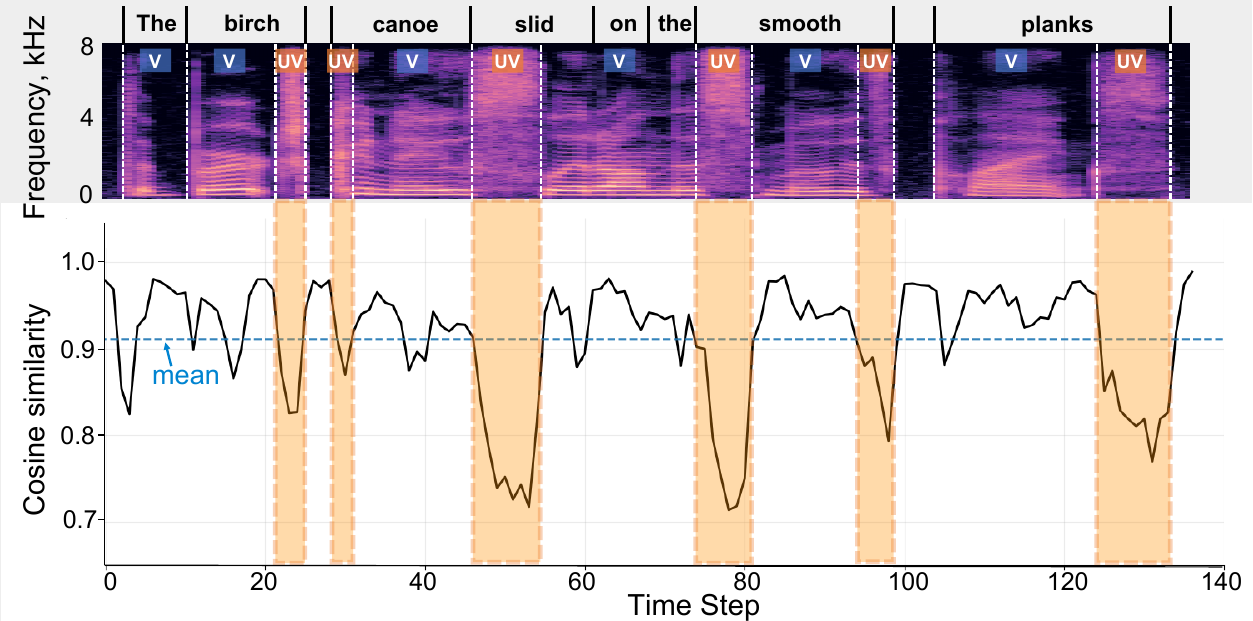}
\caption{V/UV segmentation and LR–HR embedding cosine similarity over time.
Upper: spectrogram with word boundaries and V/UV labels. Lower: cosine similarity; orange dashed regions mark UV-aligned drops, blue dashed line shows the global mean.}
\label{fig:vuv}

\end{figure}
To recover clean high-resolution (HR) speech from a low-resolution (LR) input, we propose \textbf{CodecFlow}, a bandwidth-extension framework built upon the Descript Audio Codec (DAC) \cite{kumar2023high} together with a conditional flow-matching (CFM) network for latent bandwidth conversion, as illustrated in Figure~\ref{fig:overall}. Originally proposed for high-fidelity waveform reconstruction, DAC encodes raw audio into discrete latent tokens via residual vector quantization (RVQ). Leveraging this structured discrete codec space, CodecFlow performs bandwidth extension through conditional latent transformation rather than direct waveform prediction.

Given LR speech $\mathbf{w}_l \in \mathbb{R}^{B \times N_l}$, the system first applies a voicing extractor to derive a frame-level voicing sequence $\mathbf{s} \in \{0,1,2\}^{B \times T}$, where $0$, $1$, and $2$ denote silence, unvoiced, and voiced states, respectively, which provides explicit excitation cues to better preserve voiced–unvoiced transitions during latent conversion. In parallel, the DAC encoder processes $\mathbf{w}_l$ to produce a continuous LR latent representation $\mathbf{z}_l \in \mathbb{R}^{B \times D \times T}$. We then introduce a CFM-based Flow Embedding Converter (FEC) that conditions on both $\mathbf{z}_l$ and $\mathbf{s}$ to generate the corresponding HR latent embedding $\mathbf{z}_h \in \mathbb{R}^{B \times D \times T}$. The converted latent is subsequently discretized by a modified RVQ—adapted from the original DAC quantizer—to obtain $\mathbf{z}_h^{q} \in \mathbb{R}^{B \times D \times T}$. Finally, the DAC decoder reconstructs the HR waveform $\mathbf{w}_h \in \mathbb{R}^{B \times N_h}$ from $\mathbf{z}_h^{q}$, completing the LR-to-HR bandwidth-extension pipeline. 

\subsection{Voicing Extractor}
We introduce an explicit voicing extractor to provide frame-level speech state cues for bandwidth conversion. 
Voiced and unvoiced frames exhibit substantially different high-frequency energy distributions. Moreover, as shown in Figure~\ref{fig:vuv}, we observe that LR and HR embeddings are more similar in voiced regions than in unvoiced ones. 
Incorporating voicing labels as an auxiliary condition therefore offers the model an explicit structural prior on spectral energy patterns, facilitating more accurate high-frequency reconstruction.

The voicing extractor consists of two parallel branches that operate on the input waveform $\mathbf{w}_l$. 
The first branch performs RMS-based silence detection by computing frame-wise energy in the dB domain and defining an adaptive silence threshold as the 10-th percentile energy plus a 10\,dB margin; frames below this threshold are marked as silence, yielding a binary silence mask $\mathbf{m}_s \in \{0,1\}^{B \times T}$. 
The second branch performs pitch-based voiced/unvoiced detection using Parselmouth~\cite{DBLP:journals/jphonetics/JadoulTB18}\footnote{\url{https://github.com/YannickJadoul/Parselmouth}} with an F0 range of 50--800\,Hz, where non-zero F0 values indicate voiced speech and zero values indicate unvoiced speech, producing a voiced/unvoiced sequence $\mathbf{s}_{uv} \in \{1,2\}^{B \times T}$. 
The final voicing labels are obtained via element-wise masking $\mathbf{s} = \mathbf{m}_s \odot \mathbf{s}_{uv}$, followed by length alignment to the embedding sequence and temporal refinement using a 5-frame majority-vote smoothing window.

\subsection{Flow Embedding Converter (FEC)}

To bridge low-resolution (LR) and high-resolution (HR) speech in the codec latent domain, we introduce a Flow Embedding Converter (FEC) that performs bandwidth conversion in the continuous embedding space, in contrast to prior approaches \cite{fang2025vector, liu2025neural} that directly transform quantized token sequences. We observe that despite substantial spectral differences in waveform space, LR and HR speech remain closely aligned in the continuous codec latent space, making continuous embedding conversion a natural and stable formulation for cross-resolution alignment and conditional generation. FEC is based on conditional flow matching (CFM), a continuous-time generative formulation widely adopted in recent speech enhancement \cite{lee2025flowse, jung2024flowavse} and bandwidth extension \cite{yun2025flowhigh, choi2025universr} tasks. CFM performs direct velocity regression along a predefined transport path, enabling stable and efficient conditional modeling without complex noise schedules or multi-stage training.

Given the LR codec embedding $\mathbf{z}_l$ and the frame-level voicing sequence $\mathbf{s}$, we fuse them into a unified condition $\mathbf{c}=\mathcal{F}_c(\mathbf{z}_l,\mathbf{s})$ through a lightweight convolution-based projection. The FEC then parameterizes a conditional continuous-time velocity field $v_t$ that transports a base distribution toward the HR embedding distribution under condition $\mathbf{c}$. Architecturally, the velocity network adopts a U-Net \cite{ronneberger2015u} style Conformer \cite{gulati2020conformer} backbone: the intermediate latent state and the fused condition are concatenated and projected into a hidden space, followed by symmetric encoder--decoder Conformer stacks with skip connections to jointly capture local temporal structures and long-range dependencies.

Training follows the standard CFM formulation. Let $\mathbf{z}_h^{(0)} \sim \mathcal{N}(\mathbf{0},\mathbf{I})$ be a base sample and $\mathbf{z}_h^{(1)}$ the ground-truth HR codec embedding extracted by feeding HR speech into the DAC encoder. We construct a linear transport path as in Eq.~\ref{eq:lt},
\begin{equation}
\psi_t(\mathbf{z}_h^{(0)}) = (1 - t)\mathbf{z}_h^{(0)} + t \mathbf{z}_h^{(1)}, \quad t \sim \mathcal{U}(0,1),
\label{eq:lt}
\end{equation}
whose target velocity is $\mathbf{z}_h^{(1)}-\mathbf{z}_h^{(0)}$. The model is optimized by regressing the conditional velocity field along this path as in Eq.~\ref{eq:cfm_loss},
\begin{equation}
\mathcal{L}_{\text{CFM}} =
\mathbb{E}_{t,\, p(\mathbf{z}_h^{(0)}),\, q(\mathbf{z}_h^{(1)})}
\big[
\Vert v_t(\psi_t(\mathbf{z}_h^{(0)}), \mathbf{c})
- (\mathbf{z}_h^{(1)} - \mathbf{z}_h^{(0)}) \Vert_2^2
\big],
\label{eq:cfm_loss}
\end{equation}
where $p(\cdot)$ denotes a simple base distribution and $q(\cdot)$ corresponds to the empirical data distribution in the HR embedding space. To enhance conditional controllability, we adopt classifier-free guidance (CFG) by randomly dropping the condition during training and introducing a learnable null condition. At inference, the conditional and unconditional velocity predictions are combined as in Eq.~\ref{eq:cfg},
\begin{equation}
v_{t, \text{cfg}} =
v_t(\psi_t(\mathbf{z}_h^{(0)})) +
\alpha \Big(
v_t(\psi_t(\mathbf{z}_h^{(0)}), \mathbf{c}) -
v_t(\psi_t(\mathbf{z}_h^{(0)}))
\Big),
\label{eq:cfg}
\end{equation}
where $\alpha$ is the guidance scale. We additionally apply channel-wise embedding normalization based on dataset statistics to mitigate scale mismatch between LR and HR embeddings, which improves training stability.

\subsection{Structure-Constrained Residual Vector Quantization (SC-RVQ)}
Residual Vector Quantization (RVQ) performs multi-stage quantization by progressively encoding residual errors with a stack of codebooks, yielding a hierarchical coarse-to-fine approximation. Despite this structured design, continuous codec embeddings across different bandwidths may remain closely aligned while their corresponding discrete token indices diverge due to quantization boundaries and codebook partitioning, introducing instability in downstream decoding. To mitigate this mismatch, we propose a Structure-Constrained Residual Vector Quantizer (SC-RVQ), which augments standard RVQ with lightweight structural regularization to better align continuous embeddings with their discrete counterparts.

SC-RVQ introduces two complementary objectives. First, inspired by margin-based metric learning~\cite{guo2022multi}, we enlarge the gap between the nearest and second-nearest codebook entries to sharpen quantization decision boundaries. Let the distance matrix between embeddings and codebook vectors be $\mathbf{d} \in \mathbb{R}^{B \times T \times K}$, and denote the smallest and second-smallest distances as $\mathbf{d}_1, \mathbf{d}_2 \in \mathbb{R}^{B \times T}$. The margin loss is defined in Eq.~\ref{eq:margin_loss} as 
\begin{equation}
\mathcal{L}_{\text{margin}}
= \max \bigl(0,\, \gamma - (\mathbf{d}_2 - \mathbf{d}_1)\bigr),
\label{eq:margin_loss}
\end{equation}
where $\gamma$ is a predefined margin.

To preserve hierarchical consistency across RVQ stages, we further encourage the residual energy to decay monotonically with depth. Let $\mathbf{E}_i$ denote the mean squared residual energy at layer $i$. The monotonic constraint is formulated in Eq.~\ref{eq:mono_loss} as
\begin{equation}
\mathcal{L}_{\text{mono}}
= \max \bigl(0,\, \mathbf{E}_i - \rho \mathbf{E}_{i-1}\bigr),
\label{eq:mono_loss}
\end{equation}
where $\mathbf{E}_i = \frac{1}{DT} \lVert \mathbf{z} - \mathbf{z}_i^{q} \rVert_2^2$, and $\rho \in (0,1)$ controls the allowed decay ratio. This regularization enforces a structured coarse-to-fine decomposition and stabilizes multi-stage quantization. The final objective combines the original RVQ reconstruction and commitment terms with the above regularizers, as shown in Eq.~\ref{eq:rvq_loss}, 
\begin{equation}
\mathcal{L}_{\text{SC-RVQ}}
= \mathcal{L}_{\text{RVQ}}
+ \lambda_m \mathcal{L}_{\text{margin}}
+ \lambda_r \mathcal{L}_{\text{mono}},
\label{eq:rvq_loss}
\end{equation}
where $\lambda_m$ and $\lambda_r$ are balancing coefficients.

\subsection{Training and Inference Scheme}
The model is trained in a three-stage procedure. In the first stage, we train the DAC codec augmented with the proposed SC-RVQ using the reconstruction loss $\mathcal{L}_{\text{recon}}$, the adversarial loss $\mathcal{L}_{\text{adv}}$, and the structural regularization $\mathcal{L}_{\text{SC-RVQ}}$, initialized from publicly available pretrained checkpoints\footnote{\url{https://github.com/descriptinc/descript-audio-codec}}. In the second stage, we train the CFM-based flow prediction network from the Flow Embedding Converter (FEC) with the flow-matching objective $\mathcal{L}_{\text{CFM}}$. To construct paired embedding data, both the low-resolution (LR) waveforms $\mathbf{w}_l$ and the high-resolution (HR) waveforms $\mathbf{w}_h$ are passed through the pretrained DAC encoder from stage 1 to obtain the corresponding continuous embeddings $\mathbf{z}_l$ and $\mathbf{z}_h$, which serve as supervision targets for the converter. In the final stage, we integrate the trained FEC into the full DAC pipeline for joint fine-tuning, where the FEC and SC-RVQ are kept fixed while the DAC encoder and decoder are unfrozen and optimized again with $\mathcal{L}_{\text{recon}}$ and $\mathcal{L}_{\text{adv}}$ to refine end-to-end reconstruction quality. 

During inference, we first extract the voicing sequence $\mathbf{s}$ and the LR latent embedding $\mathbf{z}_l$ from the input LR speech $\mathbf{w}_l$. These conditions are then fed into the flow prediction network together with an initialized latent state $\mathbf{z}_h^{(0)}$, where the network parameterizes a velocity field $v_t$ in the latent space. An ODE solver subsequently integrates this velocity field from $t=0$ to $t=1$ under classifier-free guidance through a sequence of discretized time steps, yielding the converted HR latent embedding $\mathbf{z}_h$. The resulting latent is then passed through the SC-RVQ to produce the quantized embedding $\mathbf{z}_h^{q}$, which is finally decoded by the DAC decoder to reconstruct the HR waveform $\mathbf{w}_h$.

\section{Experiments}
\subsection{Dataset}
We use a 40-hour LibriTTS~\cite{zen2019libritts} subset resampled to 16kHz to train the $8{\rightarrow}16\,$kHz FEC, while DAC pretraining and fine tuning uses the 100-hour subset. VCTK~\cite{yamagishi2019cstr}, resampled to 44.1kHz, is used to train the $8{\rightarrow}44.1\,$kHz FEC.
System evaluation uses 50 random TIMIT~\cite{garofolo1993darpa} utterances plus 50 utterances from four held-out VCTK speakers (p225–p228), resampled.

\subsection{Implementation Details} 
The 1024-dimensional input is concatenated with an equally sized fused conditioning vector, to which a 0.1 dropout is applied for classifier-free guidance (CFG). 
The resulting embedding is projected to 256 dimensions and combined with sinusoidal time and positional encodings. 
The sequence is then processed by a Conformer encoder–decoder architecture, where both the encoder and decoder comprise three Conformer layers with a model dimension of 256, 4 attention heads, and a feed-forward network (FFN) with dimension of 1024. 
For FEC training, we use a batch size of 32 and a learning rate of $10^{-4}$, with an early stopping patience of 5 epochs.
During inference, target latent embeddings are generated using a 25-step Euler ODE solver with a guidance scale of 1.5. 

The DAC codec is pretrained for 200k steps and the overall architecture is subsequently fine-tuned with the proposed FEC module for another 200k steps. 
During DAC pretraining, the loss weights for the margin loss $\mathcal{L}_{\rm margin}$ and mono loss $\mathcal{L}_{\rm mono}$ are set to $\lambda_{\text{m}}=0.25$ and $\lambda_{\text{r}}=0.25$, respectively. 
Training is performed using the AdamW optimizer with a learning rate of $1\times10^{-4}$ and a batch size of 16. 
All experiments are conducted on NVIDIA L40S GPUs. 
For the original DAC architectural configurations and remaining implementation details, we follow the official open-source implementation.

\subsection{Evaluation Metrics}
We evaluate reconstruction fidelity using both spectral and perceptual metrics. 
Speech spectral accuracy is evaluated via Log-Spectral Distance (LSD)~\cite{gray2003distance,yun2025flowhigh}, including low-frequency (LSD-LF) and high-frequency (LSD-HF) variants split at 4\,kHz (lower is better). Speech perceptual quality is assessed using VISQOL~\cite{chinen2020visqol}\footnote{\url{https://github.com/google/visqol}} and NISQA~\cite{mittag2021nisqa}\footnote{\url{https://github.com/gabrielmittag/NISQA}}, reporting overall MOS and Colouration (COL) sub-scores (higher is better). COL is particularly effective at capturing common BWE artifacts, such as muffled spectra.

\section{Results}
\begin{table*}[t]
\centering
\caption{Bandwidth extension evaluation at 16 kHz and 44.1 kHz targets with multiple baseline models. Best results are highlighted in \textbf{bold}, and second-best results are \underline{underlined}.}
\label{tab:baselines}
\resizebox{\linewidth}{!}{
    \begin{tabular}{lcccccccccccc}
    \toprule
    \multirow{2}{*}{\textbf{Models}} 
    & \multicolumn{6}{c}{\textbf{8 kHz $\rightarrow$ 16kHz}} 
    & \multicolumn{6}{c}{\textbf{8 kHz $\rightarrow$ 44.1 kHz}} \\
    \cmidrule(lr){2-7} \cmidrule(lr){8-13}
    & \textbf{LSD}$\downarrow$ & \textbf{LSD-LF}$\downarrow$ & \textbf{LSD-HF}$\downarrow$ & \textbf{VISQOL}$\uparrow$ & \textbf{MOS}$\uparrow$ & \textbf{COL}$\uparrow$ 
    & \textbf{LSD}$\downarrow$ & \textbf{LSD-LF}$\downarrow$ & \textbf{LSD-HF}$\downarrow$ & \textbf{VISQOL}$\uparrow$ & \textbf{MOS}$\uparrow$ & \textbf{COL}$\uparrow$ \\
    \midrule
    Input (LR Speech) & 5.36 & 1.16 & 7.49 & 2.83 & 3.77 & 3.20 & 5.50 & 0.81 & 6.07 & 2.16 & 3.85 & 3.28 \\
    Target (HR Speech) & -    & -    & -    & 4.51 & 4.29 & 4.06 & -    & -    & -    & 4.73 & 4.50 & 4.32 \\ 
    \midrule
    NU-Wave2~\cite{han2022nu}       & 1.75 & 0.51 & 2.42 & 2.60 & 3.19 & 2.84 & 1.50 & 0.72 & 1.62 & 2.19 & 3.49 & 2.98 \\
    AP-BWE~\cite{lu2024towards}        & \underline{1.24} & 0.46 & \underline{1.69} & 2.42 & \underline{3.97} & \underline{3.79} & \underline{1.14} & \textbf{0.23}& \underline{1.25} & 3.23 & \textbf{4.44} & \underline{4.25} \\
    Fre-Painter~\cite{kim2024audio}   & 1.35 &\textbf{0.34} & 1.87 & \underline{2.64} & 3.76 & 3.69 & 1.30 & \underline{0.62} & 1.40 & \underline{3.31} & 4.01 & 3.99 \\
    FlowHigh~\cite{yun2025flowhigh} & 1.34 & \underline{0.36} & 1.85 & 2.59 & 3.79 & 3.75 & 1.29 & 0.62 & 1.39 & \textbf{3.49} & 3.92 & 3.93 \\
    \rowcolor{gray!15} 
    \textbf{CodecFlow (Ours)}    & \textbf{1.01} & 0.64 & \textbf{1.27} & \textbf{2.72} & \textbf{4.25} & \textbf{4.04} & \textbf{0.93} & 0.67 & \textbf{0.98} & 3.30 & \underline{4.42} &\textbf{4.25} \\ 
    % \textbf{CodecFlow (AE)}    & \textbf{1.01} & 0.57 & \textbf{1.30} & \textbf{2.92} & \textbf{4.26} & \textbf{4.06} & \textbf{0.93} & 0.67 & \textbf{0.98} & 3.30 & \underline{4.42} &\textbf{4.25} \\ 
    \bottomrule
    \end{tabular}
}
\end{table*}

\begin{figure}[t]
\centering
\includegraphics[width=\linewidth]{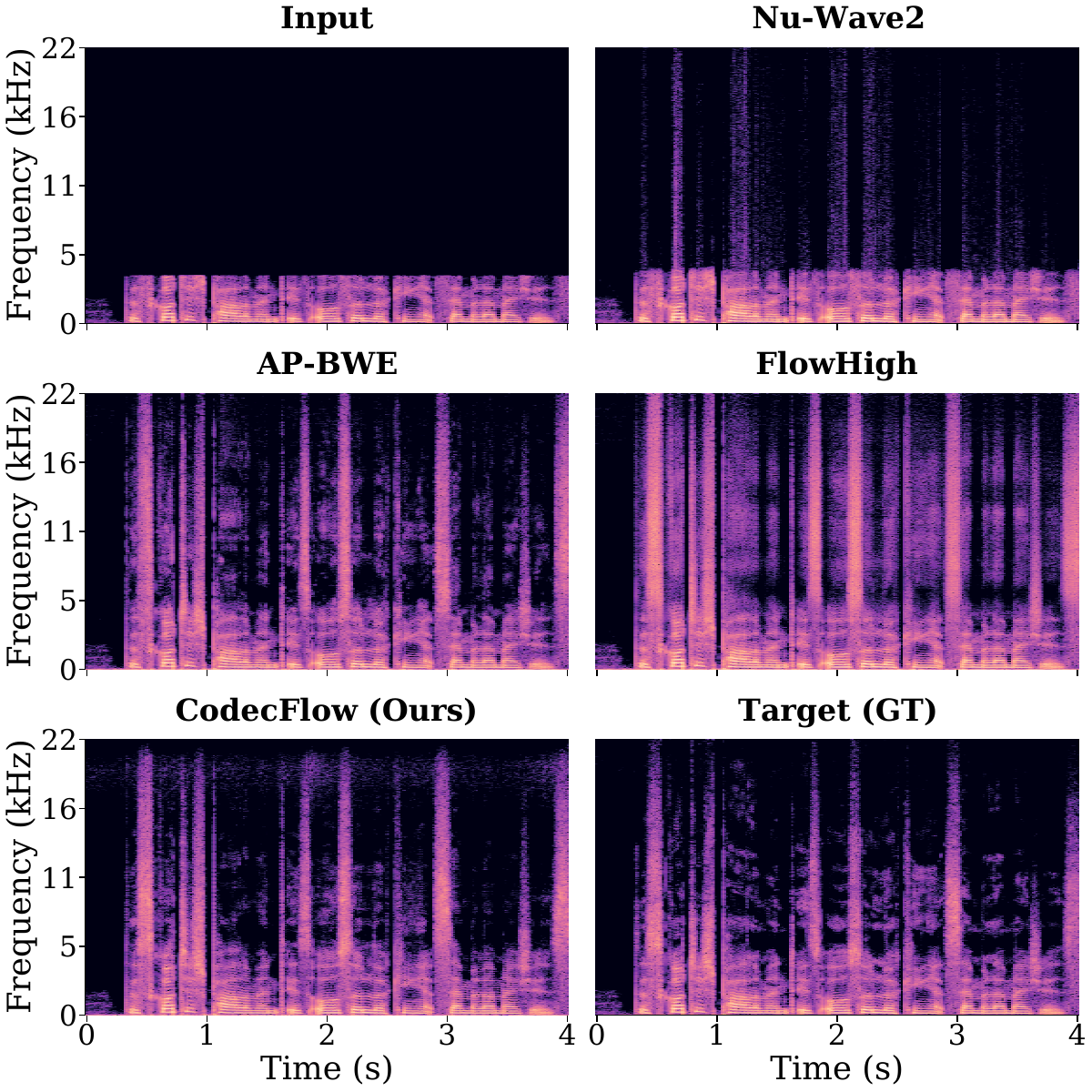}
\caption{Spectrogram comparison between the 8 kHz input, the \textbf{44.1 kHz} ground truth (GT), and model outputs, including NU-Wave2, AP-BWE, FlowHigh, and the proposed CodecFlow.}
\label{fig:mel_44}
\end{figure}

\begin{figure}[t]
\centering
\includegraphics[width=\linewidth]{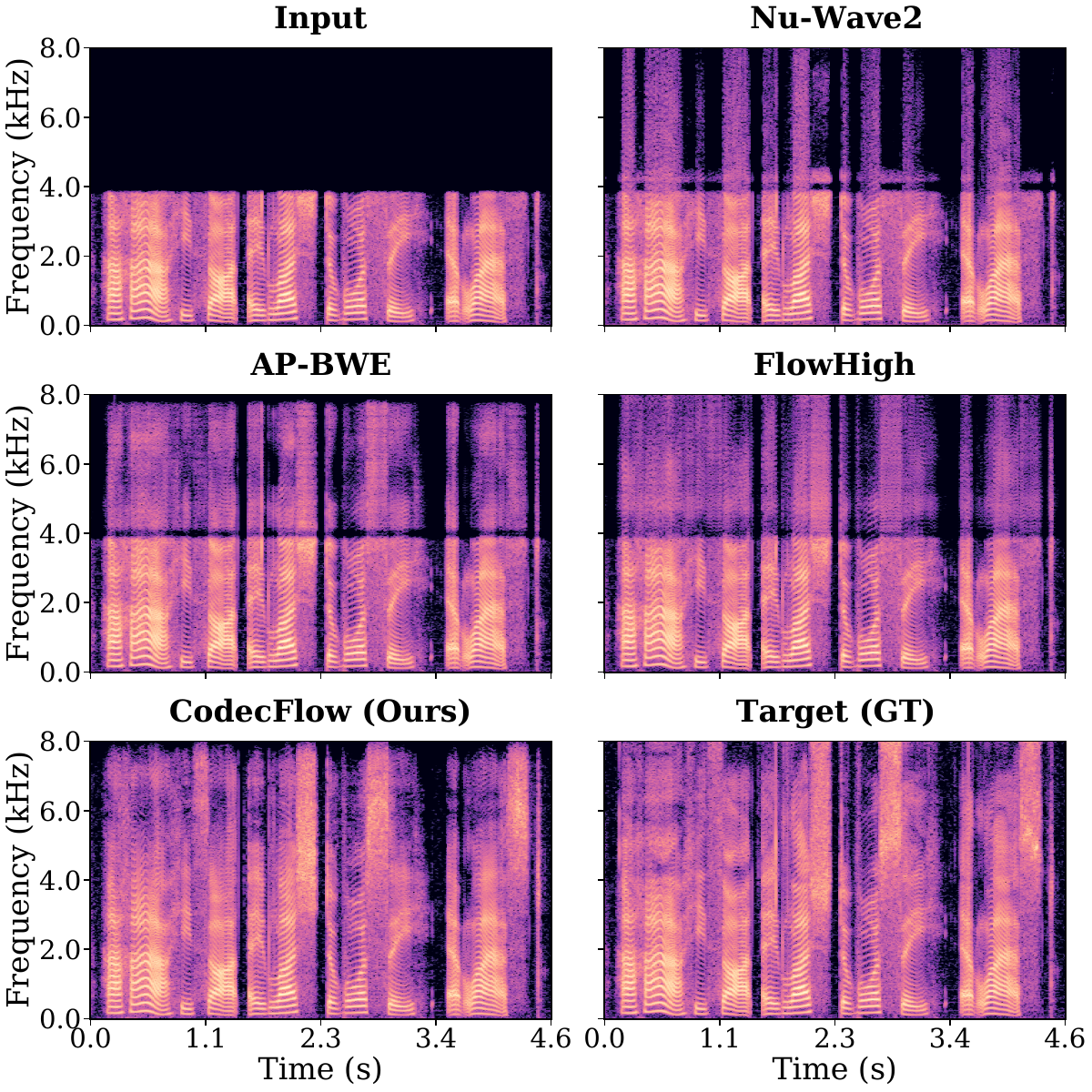}
\caption{Spectrogram comparison between the 8 kHz input, the \textbf{16 kHz} ground truth (GT), and model outputs, including NU-Wave2, AP-BWE, FlowHigh, and the proposed CodecFlow.}
\label{fig:mel_16}
\end{figure}

\subsection{Baseline Comparison}

To comprehensively evaluate CodecFlow, we compare it against representative bandwidth extension methods spanning different modeling paradigms, including Nu-Wave2~\cite{han2022nu}, AP-BWE~\cite{lu2024towards}, Fre-Painter~\cite{kim2024audio}, and CFM-based FlowHigh~\cite{yun2025flowhigh}.

Table~\ref{tab:baselines} summarizes results under two upsampling settings: 8\,kHz$\rightarrow$16\,kHz and 8\,kHz$\rightarrow$44.1\,kHz. 
For the 16\,kHz task, CodecFlow achieves the best overall performance across most objective and perceptual metrics, obtaining the lowest Log-Spectral Distance and high-frequency LSD, which indicates more accurate reconstruction of missing high-frequency components. 
In perceptual evaluation, CodecFlow records the highest VISQOL, MOS, and Coloration scores, suggesting improved clarity and reduced spectral distortion relative to competing approaches. 
These gains can be attributed to operating in a compact neural codec latent space combined with voicing-aware conditional flow modeling, which enhances high-frequency detail recovery while preserving perceptual consistency.

Under the more demanding 44.1\,kHz setting, CodecFlow continues to demonstrate strong spectral fidelity, achieving the lowest overall LSD and LSD-HF. 
Although FlowHigh reports the highest VISQOL and AP-BWE attains the highest MOS, CodecFlow achieves a comparable MOS and matches AP-BWE in Coloration. 
While certain baselines occasionally yield lower LSD-LF values, CodecFlow consistently provides superior high-frequency reconstruction and maintains competitive perceptual quality across both sampling rates, indicating stable performance under large upsampling ratios.

Figures~\ref{fig:mel_44} and \ref{fig:mel_16} present spectrogram comparisons between the low-resolution input, model outputs, and ground-truth references under the 8\,kHz $\rightarrow$ 44.1\,kHz and 8\,kHz $\rightarrow$ 16\,kHz settings, respectively.
For the 8\,kHz $\rightarrow$ 44.1\,kHz case using the VCTK utterance \textit{p228\_048\_mic1.wav}, Nu-Wave2 restores limited upper-band energy under the large upsampling ratio.
AP-BWE extends the bandwidth but introduces spectral smearing and vertical artifacts, while FlowHigh produces overly dense high-frequency textures that deviate from the reference.
In contrast, CodecFlow suppresses these artifacts and reconstructs structured components up to 22\,kHz, yielding spectral patterns more consistent with the ground truth.

Under the 8\,kHz $\rightarrow$ 16\,kHz setting using the TIMIT utterance \textit{fdac\_si2104.wav}, the emphasis shifts from bandwidth reach to spectral cleanliness.
Nu-Wave2 again shows limited restoration above 4\,kHz with fragmented harmonics, and AP-BWE exhibits visible vertical streaking and temporal smearing.
FlowHigh tends to underestimate high-frequency energy, resulting in weaker and less structured textures.
CodecFlow instead produces smoother and more continuous harmonic patterns within the 4–8\,kHz band while maintaining spectral envelopes closely aligned with the ground truth.

\begin{figure}[t]
\centering
\includegraphics[width=\linewidth]{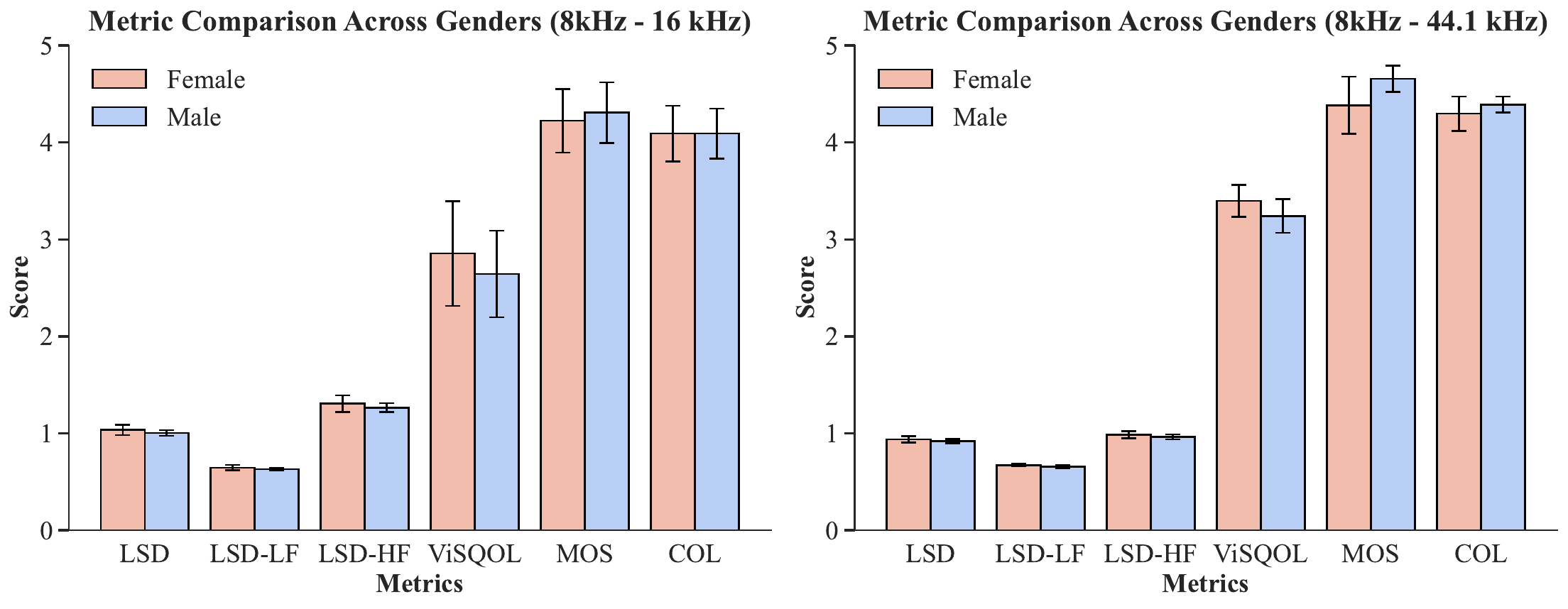}
\caption{Gender-wise comparison of objective and perceptual metrics under two bandwidth-extension settings (8\,kHz$\rightarrow$16\,kHz and 8\,kHz$\rightarrow$44.1\,kHz).}
\label{fig:gender}
\end{figure}

To assess robustness across speaker characteristics, we further analyze performance on 10 male and 10 female utterances (Fig.~\ref{fig:gender}). 
Despite the increased difficulty of high-frequency reconstruction for female voices due to higher fundamental frequencies, CodecFlow exhibits comparable performance across genders for both objective and perceptual metrics. 
The substantial overlap in error bars indicates no statistically meaningful disparity, suggesting that the proposed model maintains consistent generation quality independent of speaker gender.

% name of table 2, name of the baseline models
% model architecture of baseline models
\subsection{Ablation Study}

\begin{table}[t!]
\centering
\caption{Ablation study of converter variants in CodecFlow, where CodecReg denotes direct latent residual regression, CFM-Conf and CFM-UConf denote conditional flow–based converters with Conformer and U-Conformer backbones, “w/o FT” indicates training without fine-tuning, and CodecFlow denotes the full fine-tuned model.}
\label{tab:ablation}
\setlength{\tabcolsep}{2.5pt} 
\resizebox{\columnwidth}{!}{
\begin{tabular}{lcccccc}
\toprule
\textbf{Method} & \textbf{LSD}$\downarrow$ & \textbf{LSD-LF}$\downarrow$ & \textbf{LSD-HF}$\downarrow$ & \textbf{VISQOL}$\uparrow$ & \textbf{MOS}$\uparrow$ & \textbf{COL}$\uparrow$ \\
\midrule
\multicolumn{7}{l}{\textit{\textbf{Task: 8\,kHz $\to$ 16\,kHz}}} \\
\midrule
CodecReg (w/o FT)          & 1.19 & 0.67 & 1.53 & 2.46 & 3.77 & 3.63 \\
CFM-Conf (w/o FT)          & 1.33 & 0.72 & 1.73 & 2.24 & 3.82 & 3.70 \\
CFM-UConf (w/o FT)    & 1.21 & 0.82 & 1.48 & 2.29 & 4.05 & 3.90 \\
\textbf{CodecFlow} & \textbf{1.01} & \textbf{0.64} & \textbf{1.27} & \textbf{2.72} & \textbf{4.25} & \textbf{4.04} \\

\midrule
\multicolumn{7}{l}{\textit{\textbf{Task: 8\,kHz $\to$ 44.1\,kHz}}} \\
\midrule
CodecReg (w/o FT) &  4.86 & 1.17 & 5.34 & 1.92 & 2.67 & 2.43 \\
CFM-Conf (w/o FT) & 1.26 & 0.91 & 1.32 & 2.73 & 3.78 & 3.54  \\
CFM-UConf (w/o FT) & 1.04 & 0.72 & 1.10 & 2.87 & 3.90 & 3.73\\
\textbf{CodecFlow} & \textbf{0.93} & \textbf{0.67} & \textbf{0.98} & \textbf{3.30} & \textbf{4.42} & \textbf{4.25} \\

\bottomrule
\end{tabular}%
}
\end{table}

To evaluate the effects of different latent conversion models and training strategies in CodecFlow, we conduct ablation studies under two bandwidth extension settings (8\,kHz$\rightarrow$16\,kHz and 8\,kHz$\rightarrow$44.1\,kHz), with results summarized in Table~\ref{tab:ablation}. 

For the conversion module, we compare three variants. 
\textbf{CodecReg} adopts a Conformer-based regression model that directly predicts the latent residual 
$\Delta \mathbf{z} = \mathbf{z}_h - \mathbf{z}_l$, where $\mathbf{z}_h$ and $\mathbf{z}_l$ denote the high- and low-resolution codec embeddings, respectively. 
The predicted residual $\hat{\Delta \mathbf{z}}$ is then added to $\mathbf{z}_l$ to obtain the reconstructed high-resolution embedding $\hat{\mathbf{z}}_h$. 
\textbf{CFM-Conf} replaces direct regression with conditional flow matching (CFM) while retaining the same Conformer backbone, 
whereas \textbf{CFM-UConf} employs a U-Conformer architecture, i.e., a Conformer augmented with U-Net–style skip connections for multi-scale feature aggregation. 
All three variants share identical backbone configurations with the proposed CodecFlow and are trained without end-to-end fine-tuning.

We first compare the CFM variants with the CodecReg baseline. 
While CodecReg performs reasonably well under the 16\,kHz setting, its performance degrades substantially for the 44.1\,kHz task, indicating limited capacity in modeling the highly non-linear latent transformation required for large upsampling ratios. 
In contrast, CFM-based variants exhibit more stable spectral reconstruction and improved perceptual quality, suggesting that flow-based modeling better captures the conditional latent distribution.

Upgrading the backbone from a stacked Conformer to a U-Conformer further improves perceptual performance, highlighting the benefit of multi-scale feature propagation. 
Finally, enabling joint fine-tuning in the full CodecFlow pipeline yields consistent gains across both objective and subjective metrics, demonstrating the effectiveness of end-to-end optimization with the codec decoder.

\begin{figure}[t]
\centering
\includegraphics[width=0.5\textwidth]{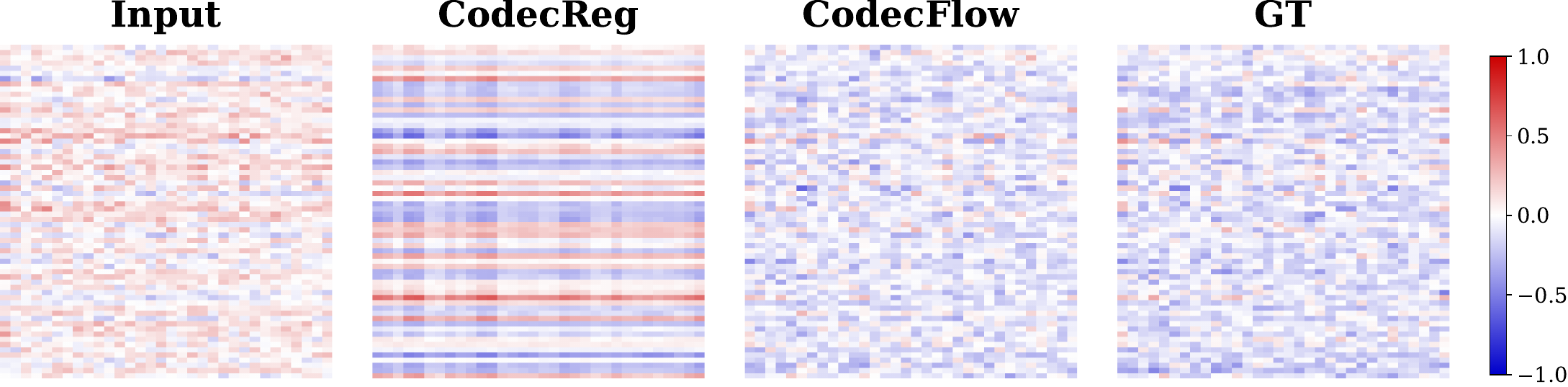}
\caption{Comparison of continuous codec latent embeddings across LR input, CodecReg, CodecFlow and HR (\textbf{16 kHz}) ground truth.}
\label{fig:emb_16}

\end{figure}
\begin{figure}[t]
\centering
\includegraphics[width=0.5\textwidth]{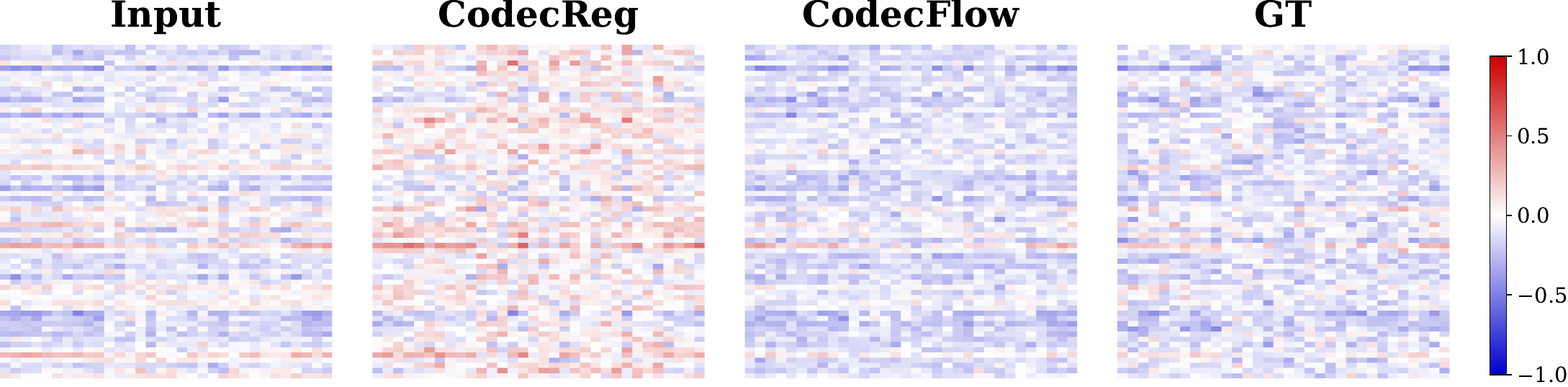}
\caption{Comparison of continuous codec latent embeddings across LR input,  CodecReg, CodecFlow, and HR (\textbf{44.1 kHz}) ground truth.}
\label{fig:emb_44}
\end{figure}

Figure \ref{fig:emb_16} and \ref{fig:emb_44} visualizes the codec latent embeddings of the LR input, the HR ground truth, and the corresponding predictions from CodecReg and CodecFlow under both 8 kHz $\rightarrow$ 16 kHz and 16 kHz $\rightarrow$ 44 kHz settings. The LR embedding exhibits relatively smooth and low-variance structures, whereas the HR ground truth displays richer spatial variations and more coherent organization across channels. CodecReg produces more scattered and locally inconsistent activations, with weaker spatial continuity and reduced channel-wise coherence compared to the ground truth, suggesting that direct latent regression does not sufficiently preserve the intrinsic structural geometry of the target codec embedding space. In contrast, CodecFlow produces embeddings that more faithfully recover the structural patterns and activation distribution of the ground truth, with improved spatial coherence and reduced irregular artifacts. This comparison indicates that modeling bandwidth conversion as flow-based latent transport better preserves the intrinsic structure of the codec space, leading to more stable and semantically consistent high-resolution representations than direct regression.

\section{Conclusion}
This work has presented CodecFlow, a neural codec–based bandwidth extension framework that reconstructs or infers high-resolution speech from low-resolution inputs within a compact latent space. 
Experiments on both $8{\rightarrow}16\,$kHz and $8{\rightarrow}\,$44.1kHz demonstrate that CodecFlow outperforms or matches current BWE models and consistently delivers strong spectral fidelity and perceptual quality while effectively suppressing high-frequency artifacts. 
Ablation studies indicate that flow-based latent modeling and joint fine-tuning are key to stable bandwidth reconstruction across resolution scales. 
Visual analyses further confirm coherent high-frequency structures and robustness across speaker characteristics.

\bibliographystyle{IEEEtran}
\bibliography{mybib}

\end{document}